\documentclass[preprint,11pt,5p,times,number,sort&compress]{elsarticle}

\usepackage{amssymb}
\usepackage{amsmath}
\usepackage{bm}
\usepackage{subfigure}
\usepackage[colorlinks=true, linkcolor=blue, citecolor=blue, urlcolor=blue]{hyperref}

\usepackage[final]{microtype}
\journal{Physics Letters B}

\begin{document}

\begin{frontmatter}

\title{Low-energy scattering of the $J/\psi \pi$ and $J/\psi K$ systems} 

\author[lab1]{Jiang Yan}
\ead{yjiang@itp.ac.cn}
\author[lab1]{Xiong-Hui Cao}
\ead{xhcao@itp.ac.cn}
\author[lab2]{Meng-Lin Du}
\ead{du.ml@uestc.edu.cn}
\author[lab1,lab3]{Feng-Kun Guo}
\ead{fkguo@itp.ac.cn}

\affiliation[lab1]{organization={Institute of Theoretical Physics, Chinese Academy of Sciences}, 
        city={Beijing},
        postcode={100190}, 
        country={China}}

\affiliation[lab2]{organization={School of Physics, University of Electronic Science and Technology of China},
        city={Chengdu},
        postcode={611731}, 
        country={China}}

\affiliation[lab3]{organization={School of Physical Sciences, University of Chinese Academy of Sciences},
	city={Beijing},
	postcode={100049}, 
	country={China}}    

\begin{abstract}
We investigate the low-energy interactions between the charmonium state $J/\psi$ and the light pseudoscalar mesons ($\pi$ and $K$) within the framework of dispersion relations. We demonstrate that the symmetry-breaking terms in the chiral Lagrangian induce mixing between the bare charmonium fields, necessitating a diagonalization procedure to correctly identify the physical $J/\psi$ and $\psi'$ states. Using the resulting diagonalized Lagrangian, we construct the crossed-channel amplitudes for $J/\psi J/\psi \to {\cal P}\bar{\cal P}$ and incorporate the $\pi\pi$ and $K\bar{K}$ rescattering effects through dispersion relations. This framework is used consistently both in the phenomenological extraction of the transition parameters from $\psi' \to J/\psi\pi\pi$ and in the continuation of the crossed amplitudes to the near-threshold $J/\psi{\cal P}~({\cal P}=\pi,K)$ region. As a result, we determine both the scattering lengths and the effective ranges. We obtain the upper-bound estimates $a_{J/\psi\pi}\lesssim -0.0037$~fm and $a_{J/\psi K}\lesssim -0.049$~fm, where the negative sign indicates an attractive interaction without a bound state in our convention. Our results show that the $J/\psi K$ interaction is moderately enhanced relative to the pion channel, driven by explicit chiral symmetry breaking. Furthermore, a quantitative comparison of the coupled-channel mechanism, where $J/\psi\pi$ and $J/\psi K$ couple to open-charm channels, reveals that both $J/\psi\pi$ and $J/\psi K$ scatterings are predominantly governed by the soft-gluon exchange mechanism.
\end{abstract}


\end{frontmatter}

\section{Introduction}
\label{sec1}

The low-energy interactions between heavy quarkonia and light hadrons, such as pions, kaons, and nucleons, have long attracted considerable theoretical interest~\cite{Peskin:1979va, Bhanot:1979vb}. These processes provide a unique probe for exploring the gluonic degrees of freedom inside hadrons and for elucidating the underlying mechanisms of inter-hadron interactions. In particular, they offer direct access to nonperturbative aspects of quantum chromodynamics~(QCD) that are difficult to explore through reactions involving only light hadrons. Moreover, the interaction between the $J/\psi$ and light hadrons plays a pivotal role in the phenomenology of $J/\psi$ suppression in relativistic heavy-ion collisions, which is widely regarded as a key signature for the formation of the quark-gluon plasma~(QGP)~\cite{Barnes:2003vt}. Despite decades of intensive study, a quantitative and unified understanding of these interactions remains a challenge.

A distinctive feature of heavy-quarkonium--light-hadron scattering is that the two hadrons share no common valence quark constituents. Consequently, these processes are suppressed by the Okubo--Zweig--Iizuka~(OZI) rule~\cite{Okubo:1963fa, Zweig:1964jf, Iizuka:1966fk}, and quark-exchange mechanisms, which dominate many light-hadron interactions, are absent. The interaction is therefore mediated predominantly by gluonic dynamics, making heavy quarkonia particularly clean probes of multi-gluon exchange effects and long-range gluon-induced forces in QCD.

In general, OZI-suppressed quarkonium--hadron interactions can be classified into two distinct mechanisms. The first is the coupled-channel mechanism, in which the quarkonium--hadron system couples to intermediate states containing open-charm mesons, such as $D\bar{D}^{*}$ or $D^{*}\bar{D}_s/D\bar{D}_s^*$, and subsequently rescatters back to the original channel. This mechanism may become particularly important when the open-charm thresholds lie close to the quarkonium--hadron threshold and can be further enhanced by the presence of near-threshold exotic states. The second mechanism involves multi-gluon exchange, which can be viewed as a color van der Waals interaction~\cite{Brodsky:1989jd, Brodsky:1997gh} between two color-singlet hadrons. In the heavy-quark limit, this interaction is commonly treated within the framework of the QCD multipole expansion~\cite{Gottfried:1977gp, Voloshin:1978hc, Peskin:1979va, Bhanot:1979vb, Mannel:1995jt, Luke:1992tm, Sibirtsev:2005ex, Kaidalov:1992hd, deTeramond:1997ny, Beane:2014sda, Krein:2020yor}, in which the quarkonium acts as a compact color dipole interacting with the gluon fields generated by the light hadron.

The relative importance of these two mechanisms depends sensitively on both the quantum numbers of the light hadron and the kinematic regime under consideration. In the case of the pion, additional constraints arise from chiral symmetry. As a pseudo-Nambu--Goldstone (pNG) boson of spontaneous chiral symmetry breaking of QCD, the pion decouples from hadrons in the chiral limit. 
Chiral effective theory predicts that the leading-order $S$-wave scattering amplitude between an isoscalar heavy quarkonium state, such as the $J/\psi$, and a pion vanishes, with the first nonzero contribution appearing only at ${\cal O}(m_{\pi}^2)$~\cite{Chen:1997zza} (the chiral Lagrangian for the $\psi'\to J/\psi \pi\pi$ was constructed in Ref.~\cite{Brown:1975dz} immediately after the discovery of $J/\psi$ and $\psi'$). 
Consequently, the $J/\psi \pi$ interaction is expected to be extremely weak near threshold. This qualitative expectation has been corroborated by lattice QCD calculations~\cite{Yokokawa:2006td, Liu:2008rza}, which find that the corresponding scattering length $|a_{0}|$ is of order $10^{-2}$~fm, consistent with a weakly attractive interaction. Furthermore, an approximate upper bound of $|a_{0}| \lesssim 0.02$~fm was estimated in Ref.~\cite{Liu:2012dv}.

The situation is somewhat different for the $J/\psi K$ system. Although the kaon also belongs to the pseudoscalar meson octet, the explicit breaking of ${\rm SU}(3)_f$ symmetry by the sizable strange quark mass relaxes the chiral constraints that apply to kaons.
Consequently, the chiral suppression governing the $J/\psi \pi$ interaction is partially lifted in the kaon channel. One may therefore expect the $J/\psi K$ scattering length to be sizeably larger than that of $J/\psi \pi$, while still remaining smaller than typical quarkonium--nucleon interactions. Furthermore, the proximity of open-charm strange thresholds suggests that coupled-channel effects could play a more prominent role in the $J/\psi K$ system, potentially leading to nontrivial dynamics that are more strongly suppressed in the pion channel.

Despite its clear theoretical interest, a systematic and quantitative study of low-energy $J/\psi K$ scattering remains lacking. Most existing investigations have focused either on quarkonium--nucleon interactions or on the pion channel, while the kaon case has received comparatively little attention. 
In Ref.~\cite{Wu:2024xwy}, we found that $J/\psi$--nucleon scattering in the near-threshold region is dominated by multi-gluon exchange rather than the coupled-channel mechanism. Whether the same conclusion holds for $J/\psi\pi$ and $J/\psi K$ scattering remains to be investigated.

In this work, we present a detailed study of low-energy $S$-wave scattering of $\pi$ and $K$ off $J/\psi$. Our analysis treats multi-gluon exchange and coupled-channel dynamics separately, enabling us to quantify their relative contributions to scattering lengths, effective ranges, and more generally the near-threshold amplitudes. This paper is organized as follows. In Sec.~\ref{sec2}, we discuss in detail the effects of explicit chiral symmetry-breaking terms in the chiral Lagrangian and demonstrate that diagonalizing the $\psi'$ and $J/\psi$ mass terms does not eliminate their contribution to the $\psi'\to J/\psi \pi\pi$ contact term, thereby correcting a statement in Ref.~\cite{Chen:2015jgl}. In Sec.~\ref{sec3}, we present the calculation of the near-threshold $J/\psi {\cal P}$ amplitudes and the corresponding scattering parameters using dispersion relations.
Finally, Sec.~\ref{sec4} provides a summary.

\section{Interaction between charmonia and light hadrons}
\label{sec2}

The interaction between charmonium states and light pseudoscalar mesons can be described by the following chiral effective Lagrangian~\cite{Mannel:1995jt, Chen:2015jgl},
\begin{align}\label{Lagrangian1}
	{\cal L} =&\, \operatorname{Tr_{s}}[J_{\beta}^{\dagger} J_{\alpha}] \bigg[ \frac{c_{1}^{(\alpha\beta)}}{2} \operatorname{Tr_{f}}[u_{\mu}u^{\mu}] + \frac{c_{2}^{(\alpha\beta)}}{2} \operatorname{Tr_{f}}[u_{\mu}u_{\nu}]v^{\mu}v^{\nu} \notag\\
	&\, + \frac{c_{m}^{(\alpha\beta)}}{2} \operatorname{Tr_{f}}[\chi_{+}] \bigg] + {\rm H.c.},
\end{align}
where $\alpha$ and $\beta$ label the charmonium states, and $\operatorname{Tr_{s}}[...]$ and $\operatorname{Tr_{f}}[...]$ denote traces over the spinor and flavor spaces, respectively. The composite field $J_{\alpha} = \vec{\psi}_{\alpha}\cdot \vec{\sigma} + \eta_{\alpha}$ encodes the charmonium degrees of freedom, with $\vec{\sigma}$ representing the Pauli matrices in spinor space. In Eq.~\eqref{Lagrangian1}, $v^{\mu}=(1,\vec{0})$ is the four-velocity of the charmonia in the rest frame, $\chi_{\pm} = u^{\dagger}\chi u^{\dagger} \pm u\chi^{\dagger}u$, and $\chi = 2B_{0}{\cal M}$, where $B_{0}$ is a constant related to the quark condensate in the chiral limit and ${\cal M}$ denotes the light-quark mass matrix. In the isospin limit, the pion and kaon masses squared are given by $m_{\pi}^{2} = 2B_{0}\hat{m}$ and $m_{K}^{2} = B_{0}(\hat{m}+m_{s})$, with $\hat{m}$ being the average mass of the $u$ and $d$ quarks. Furthermore, the chiral vielbein is defined as $u_{\mu} = i(u^{\dagger}\partial_{\mu}u - u\partial_{\mu}u^{\dagger})$, where 
\begin{align}
	u &= \exp\left(\frac{i\Phi}{\sqrt{2}F_{\pi}}\right),\\
	\Phi &= \begin{pmatrix}
		\dfrac{\pi^{0}}{\sqrt{2}}+\dfrac{\eta}{\sqrt{6}}&\pi^{+}&K^{+}\\
		\pi^{-}&-\dfrac{\pi^{0}}{\sqrt{2}}+\dfrac{\eta}{\sqrt{6}}&K^{0}\\
		K^{-}&\bar{K}^{0}&-\dfrac{2\eta}{\sqrt{6}}
	\end{pmatrix},
\end{align}
with $F_{\pi}$ denoting the pion decay constant in the chiral limit. After evaluating the spinor trace $\operatorname{Tr_{s}}[...]$ and focusing on the vector charmonium sector, the chiral Lagrangian in Eq.~\eqref{Lagrangian1} can be expanded in powers of the pNG fields. Up to quadratic order, this yields 
\begin{align}\label{Lagrangian2}
	{\cal L}_{\psi} =&\,  2\left(\vec{\psi}_{\beta}^{\dagger}\cdot \vec{\psi}_{\alpha} \right) \Bigg[ \frac{c_{1}^{(\alpha\beta)}}{F_{\pi}^{2}} \operatorname{Tr_{f}}[\partial_{\mu}\Phi\partial^{\mu}\Phi]\notag\\
	&\, + \frac{c_{2}^{(\alpha\beta)}}{F_{\pi}^{2}} \operatorname{Tr_{f}}[\partial_{\mu}\Phi\partial_{\nu}\Phi]v^{\mu}v^{\nu} + 2B_{0}c_{m}^{(\alpha\beta)} \tilde{m}\notag\\
	&\, - \frac{c_{m}^{(\alpha\beta)}}{F_{\pi}^{2}} 2B_{0} \operatorname{Tr_{f}}[{\cal M}\Phi^{2}] \Bigg] + {\rm H.c.},
\end{align}
where $\tilde{m} = \operatorname{Tr_{f}}[{\cal M}]$. In the isospin limit, one has $B_{0}\tilde{m} = m_{\pi}^{2}/2 + m_{K}^{2}$.

The inclusion of the symmetry-breaking $c_{m}$-term has been a source of theoretical ambiguity~\cite{Chen:2015jgl, Chen:2019hmz, Dong:2021lkh, Wu:2024xwy}. On the one hand, introducing this term induces mixing between charmonium fields with different $\alpha$, giving rise to the physical $J/\psi$ and $\psi'$ states; on the other hand, omitting it would eliminate all higher-order $J/\psi \psi' \pi \pi$ vertices that arise from explicit chiral symmetry-breaking operators in powers of the light quark masses. We demonstrate below that a proper diagonalization procedure resolves this dilemma. To identify the physical states, we introduce a rotation matrix ${\rm U}$,
\begin{align}
	\Psi^{(\rm phys)} &\equiv 
	\begin{pmatrix}
		\vec{\psi}_{1}^{(\rm phys.)}\\
		\vec{\psi}_{2}^{(\rm phys.)}
	\end{pmatrix} = {\rm U}^{\rm T}\Psi, \notag \\
	{\rm U} &\equiv \begin{pmatrix}
		\cos\theta_{\rm mix}&-\sin\theta_{\rm mix}\\
		\sin\theta_{\rm mix}&\cos\theta_{\rm mix}
	\end{pmatrix},\quad \Psi \equiv 
	\begin{pmatrix}
	\vec{\psi}_{1}\\
	\vec{\psi}_{2}
	\end{pmatrix}
\end{align}
where $\vec{\psi}_{1}^{(\rm phys.)}$ and $\vec{\psi}_{2}^{(\rm phys.)}$ represent the physical $J/\psi$ and $\psi'$ charmonium fields, respectively, and $\theta_{\rm mix}$ denotes the mixing angle. Collecting the quadratic terms in the charmonium fields, the mass sector of the Lagrangian can be written as ${\cal L}_{m} = \frac{1}{2}\Psi^{\dagger} {\bf M}^{2} \Psi$. The mass-squared matrix ${\bf M}^{2}$, which incorporates the explicit chiral symmetry-breaking $c_{m}^{(\alpha\beta)}$ terms, takes the form 
\begin{align}
	{\bf M}^{2} =
	\begin{pmatrix}
		m_{1}^{2}+8B_{0}\tilde{m}c_{m}^{(11)}&8B_{0}\tilde{m}c_{m}^{(12)}\\
		8B_{0}\tilde{m}c_{m}^{(12)}&m_{2}^{2}+8B_{0}\tilde{m}c_{m}^{(22)}
	\end{pmatrix},
\end{align}
where $m_{1}$ and $m_{2}$ denote the bare masses of the basis fields $\vec{\psi}_{1}$ and $\vec{\psi}_{2}$, respectively, in the absence of the symmetry-breaking $c_{m}^{(\alpha\beta)}$ terms. They can be regarded as the $1S$ and $2S$ vector-charmonium masses in the chiral limit.

Diagonalization of the mass-squared matrix ${\bf M}^{2}$ requires ${\rm U}{\bf M}^{2}{\rm U}^{\rm T}  = \operatorname{diag}(M_{J/\psi}^{2},M_{\psi'}^{2})$. This condition yields the mixing angle relation
\begin{align}
	\tan(2\theta_{\rm mix}) = \frac{16B_{0}\tilde{m}c_{m}^{(12)}}{m_{2}^{2}-m_{1}^{2} + 8B_{0}\tilde{m}\left(c_{m}^{(22)}-c_{m}^{(11)}\right)},
\end{align}
and the diagonalized Lagrangian
\begin{align}\label{Ldiag}
	{\cal L}_{\psi}^{(\rm diag.)} =&\, \frac{2}{F_{\pi}^{2}}\left(\vec{\psi}_{\beta}^{{(\rm phys.)}\dagger}\cdot \vec{\psi}_{\alpha}^{(\rm phys.)}\right) \bigg[ \tilde{c}_{1}^{(\alpha\beta)} \operatorname{Tr_{f}}[\partial_{\mu}\Phi\partial^{\mu}\Phi] \notag\\
	&\, + \tilde{c}_{2}^{(\alpha\beta)} \operatorname{Tr_{f}}[\partial_{\mu}\Phi\partial_{\nu}\Phi]v^{\mu}v^{\nu} \notag\\
	&\, - 2 \tilde{c}_{m}^{(\alpha\beta)} B_{0} \operatorname{Tr_{f}}[{\cal M}\Phi^{2}] \bigg] + {\rm H.c.},
\end{align}
where the new coefficients $\tilde{c}_{i}^{(\alpha\beta)}\ (i=1,2,m)$ are related to the original coefficients $c_{i}^{(\alpha\beta)}\ (i=1,2,m)$ via
\begin{align}
	\tilde{c}_{i}^{(11)} =&\, \frac{1}{2}\left(c_{i}^{(11)} + c_{i}^{(22)}\right) - \frac{1}{2} \cos(2\theta_{\rm mix})\left(c_{i}^{(22)} - c_{i}^{(11)}\right) \notag\\
	&\, - \sin(2\theta_{\rm mix}) c_{i}^{(12)},\\
	\tilde{c}_{i}^{(22)} =&\, \frac{1}{2}\left(c_{i}^{(11)} + c_{i}^{(22)}\right) + \frac{1}{2} \cos(2\theta_{\rm mix})\left(c_{i}^{(22)} - c_{i}^{(11)}\right) \notag\\
	&\, + \sin(2\theta_{\rm mix}) c_{i}^{(12)},\\
	\tilde{c}_{i}^{(12)} =&\, \cos(2\theta_{\rm mix}) c_{i}^{(12)} - \frac{1}{2} \sin(2\theta_{\rm mix}) \left(c_{i}^{(22)} - c_{i}^{(11)}\right).
\end{align}
Notably, for the mass-related coupling $\tilde{c}_{m}^{(\alpha\beta)}$, one obtains the explicit expressions
\begin{align}
	\tilde{c}_{m}^{(12)} &= \frac{1}{8B_{0}\tilde{m}} \Delta_{m}\sin(2\theta_{\rm mix}),\label{cm12}\\
	\tilde{c}_{m}^{(11)} &= \frac{1}{8B_{0}\tilde{m}} \bigg[ M_{J/\psi}^{2} - \Sigma_{m} + \Delta_{m} \cos(2\theta_{\rm mix}) \bigg],\label{cm11}\\
	\tilde{c}_{m}^{(22)} &= \frac{1}{8B_{0}\tilde{m}} \bigg[ M_{\psi'}^{2} - \Sigma_{m} - \Delta_{m} \cos(2\theta_{\rm mix}) \bigg],\label{cm22}
\end{align}
where $\Delta_{m} = (m_{2}^{2} - m_{1}^{2})/2$ and $\Sigma_{m} = (m_{2}^{2} + m_{1}^{2})/2$ are linear combinations of the charmonium masses squared in the chiral limit. They are introduced here only to make the diagonalization transparent and are not used as independent numerical inputs in the calculation below.

As shown in Eq.~\eqref{Ldiag}, the diagonalization procedure effectively eliminates unphysical state mixing while preserving the contact interactions in the physical basis.\footnote{The symmetry-breaking-induced contact term vanishes only if $m_{1}^{2}=m_{2}^{2}$ or $c_{m}^{(12)}=0$, as can be seen from 
$$C_m\equiv 
\begin{pmatrix}
c_m^{(11)} & c_m^{(12)}\\
c_m^{(12)} & c_m^{(22)}
\end{pmatrix}, \ 
[\mathbf M^2,C_m]=(m_1^2-m_2^2)c_m^{(12)}
\begin{pmatrix}
0 & 1\\
-1 & 0
\end{pmatrix}.
$$
} 
Consequently, the claim in footnote~2 of Ref.~\cite{Chen:2015jgl} that the $c_m$ term is removed by diagonalization requires correction. 
The transition coefficient $\tilde{c}_{m}^{(12)}$ can be constrained experimentally through the decay width of $\psi' \to J/\psi \pi \pi$.

\section{Low Energy Scattering of $J/\psi {\cal P}$}
\label{sec3}

\subsection{Multi-gluon exchange mechanism}

To obtain the $J/\psi {\cal P}$ (${\cal P}=\pi$ or $K$) scattering parameters, we evaluate the $S$-wave scattering amplitude near the corresponding threshold. As a first step, we derive the tree-level amplitude for the crossed-channel processes $J/\psi J/\psi \to \bar{\cal P}{\cal P}$ using the diagonalized chiral Lagrangian in Eq.~\eqref{Ldiag}. The resulting tree-level amplitude contains only $S$- and $D$-wave components. We write it as
\begin{align}\label{tree-amp}
	{\cal A}_{{\cal P}}(s,\cos\theta_{s}) = {\cal A}_{{\cal P},0}(s) + {\cal A}_{{\cal P},2}(s) P_{2}(\cos\theta_{s}).
\end{align}
where $P_{2}(\cos\theta_{s}) = (3\cos^{2}\theta_{s}-1)/2$ is the second-order Legendre polynomial, $s$ is the Mandelstam variable (the squared invariant mass) of the $J/\psi J/\psi$ (or ${\cal P} \bar{\cal P}$) system, and $\theta_{s}$ denotes the scattering angle in the $s$-channel center-of-mass frame. 
Note that in our normalization the factor $(2L+1)$ has been included in the definition of
the partial-wave amplitude ${\cal A}_{{\cal P},L}(s)$.
The corresponding $S$- and $D$-wave amplitudes in this normalization are
\begin{align}
	&{\cal A}_{{\cal P},0}(s) = -\frac{2 \xi_{\cal P}}{F_{\pi}^{2}} \bigg\{\tilde{c}_{1}^{(11)}\left(s-2m_{\cal P}^{2}\right) + 2\tilde{c}_{m}^{(11)}m_{\cal P}^{2}\notag\\
	&\quad + \frac{\tilde{c}_{2}^{(11)}}{12M_{J/\psi}^{2}}\left[s^{2} + 2s \left(m_{\cal P}^{2} + M_{J/\psi}^{2}\right) - 8 m_{\cal P}^{2} M_{J/\psi}^{2} \right] \bigg\},\label{S-amp}\\
	&{\cal A}_{{\cal P},2}(s) = \frac{8\xi_{\cal P}}{3 F_{\pi}^{2}} \frac{\tilde{c}_{2}^{(11)}}{M_{J/\psi}^{2}}\left(\frac{s}{4}-m_{\cal P}^{2}\right)\left(\frac{s}{4}-M_{J/\psi}^{2}\right),\label{eq:D-amp}
\end{align}
where $\xi_{\pi} = \sqrt{3/2}$ and $\xi_{K} = \sqrt{2}$ are the corresponding isospin factors. The low-energy constants~(LECs) $\tilde{c}_{1,2,m}^{(AB)}$ can be related to the charmonium chromo-polarizabilities $\alpha_{AB}$ as~\cite{Novikov:1980fa, Voloshin:2007dx, Brambilla:2015rqa, Pineda:2019mhw}
\begin{align}
	\tilde{c}_{1}^{(AB)} &= - \frac{\pi^{2}\sqrt{M_{A}M_{B}}F_{\pi}^{2}}{\beta_{0}} \alpha_{AB} \left(4 + 3 \kappa \right),\\
	\tilde{c}_{2}^{(AB)} &= \frac{12\pi^{2}\sqrt{M_{A}M_{B}}F_{\pi}^{2}}{\beta_{0}} \alpha_{AB} \kappa,\\
	\tilde{c}_{m}^{(AB)} &= - \frac{6\pi^{2}\sqrt{M_{A}M_{B}}F_{\pi}^{2}}{\beta_{0}} \alpha_{AB},
\end{align}
where $\beta_{0} = 9$ is the leading coefficient of the QCD $\beta$ function for three light flavors. 
The dimensionless parameter $\kappa$ characterizes the gluonic structure of the pion~\cite{Novikov:1980fa, Voloshin:2007dx} and needs to be determined phenomenologically.

The chromo-polarizabilities of $J/\psi$ and $\psi'$ have been investigated using various approaches; however, they remain poorly constrained. In the heavy-quark limit and the large-$N_{C}$ approximation (with $N_{C}$ being the number of colors), they were estimated to be $\alpha_{11} \approx 0.2~{\rm GeV}^{-3}$ and $\alpha_{22} \approx 12~{\rm GeV}^{-3}$~\cite{Peskin:1979va, Bhanot:1979vb, Eides:2015dtr, Eides:2017xnt}. Combining quarkonium--nucleon effective field theory~\cite{TarrusCastella:2018php} with lattice QCD data~\cite{Kawanai:2010ev} yielded $\alpha_{11} \approx 0.24~{\rm GeV}^{-3}$~\cite{TarrusCastella:2018php, Brambilla:2015rqa}. More recently, an extraction based on lattice QCD calculations~\cite{Sugiura:2017vks} within the effective potential formalism~\cite{Voloshin:1979uv} produced a significantly larger value, $\alpha_{11} = (1.6 \pm 0.8)~{\rm GeV}^{-3}$~\cite{Polyakov:2018aey}.
Nevertheless, the chromo-polarizabilities $\alpha_{AB}$ must satisfy the Cauchy--Schwarz inequality~\cite{Sibirtsev:2005ex}
\begin{align}
	\alpha_{11}\alpha_{22} \ge |\alpha_{12}|^{2}, \label{eq:inequality}
\end{align}
which imposes a nontrivial constraint on the diagonal ones. 

The off-diagonal chromo-polarizability can be extracted from the dipion transition $\psi'\to J/\psi\pi\pi$. The importance of the $S$-wave $\pi\pi$ final-state interaction (FSI) in this extraction was first noted in Ref.~\cite{Guo:2004dt}, and its impact on the extraction of the off-diagonal chromo-polarizability was subsequently discussed in Ref.~\cite{Guo:2006ya}. 
An updated phenomenological analysis of the $\pi\pi$ invariant-mass spectrum and the helicity-angular distribution in the $\psi' \to J/\psi \pi^{+}\pi^{-}$ transition~\cite{ATLAS:2016kwu, BES:2006eer}, incorporating the $c_m$ term, yields $\kappa = 0.26 \pm 0.01 \pm 0.01$ and $|\alpha_{12}| = (1.18\pm 0.01 \pm 0.05)~{\rm GeV}^{-3}$~\cite{Wu:2024xwy}. 

In this work, we update the fit to the $\pi\pi$ invariant-mass distribution and the helicity-angle distribution in $\psi'\to J/\psi\pi^+\pi^-$ by including the ATLAS~\cite{ATLAS:2016kwu}, BES~\cite{BES:1999guu}, BESII~\cite{BES:2006eer}, and the recent BESIII~\cite{BESIII:2025ozb} data. The minimization is performed with the \texttt{MINUIT} algorithm~\cite{James:1975dr, iminuit}, yielding $\kappa=0.251$ and $|\alpha_{12}|=1.107~{\rm GeV}^{-3}$. The fit uncertainties are smaller than the last quoted digits and are thus omitted.

In addition to the Cauchy--Schwarz inequality in Eq.~\eqref{eq:inequality}, Ref.~\cite{Dong:2021lkh} argued---and demonstrated explicitly using quark model wave functions---that the diagonal chromo-polarizability $\alpha_{11}$ should exceed the off-diagonal one $\alpha_{12}$, because the wave function overlap between two $J/\psi$ states following gluon emission is larger than that between a $\psi'$ and a $J/\psi$. Accordingly, we adopt $\alpha_{11} \gtrsim |\alpha_{12}| \approx 1.107~{\rm GeV}^{-3}$ in the numerical analysis below. Since the tree-level scattering length scales as $a_{J/\psi{\cal P}}\propto -\alpha_{11}$, the lower bound on $\alpha_{11}$ translates into an upper bound on $a_{J/\psi{\cal P}}$ (hence the symbol $\lesssim$ in our numerical results).

\begin{figure*}[htb]
	\centering
	\includegraphics[width=0.48\textwidth]{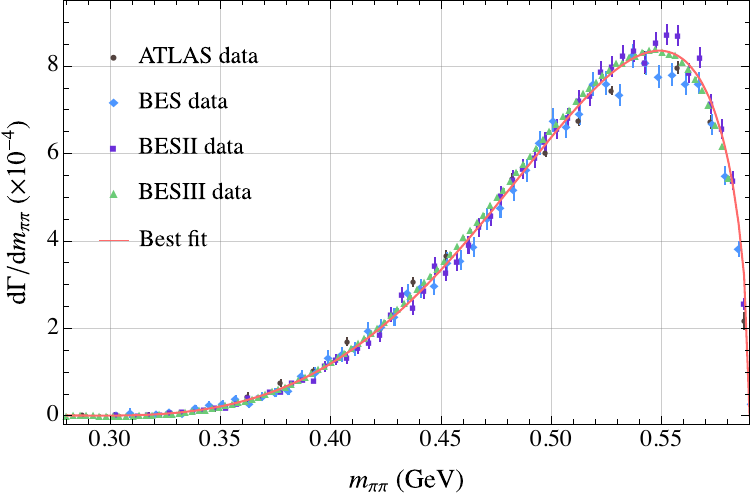}
	\includegraphics[width=0.48\textwidth]{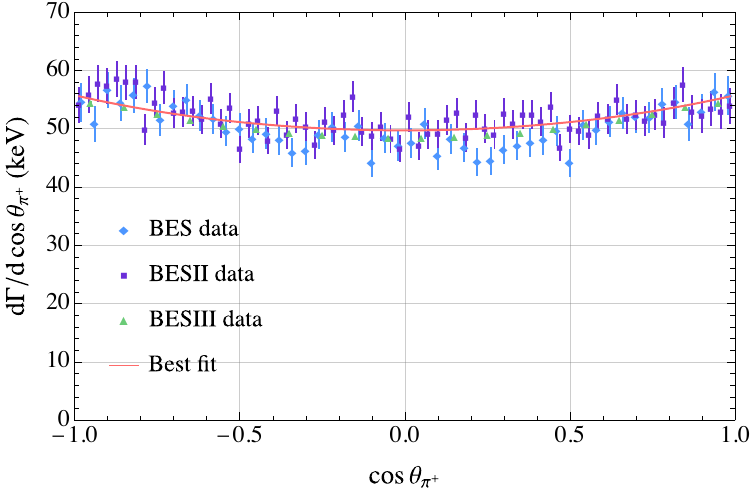}
	\caption{Fit to the ATLAS~\cite{ATLAS:2016kwu}, BES~\cite{BES:1999guu}, BESII~\cite{BES:2006eer}, and BESIII~\cite{BESIII:2025ozb} data for the dipion invariant-mass distribution (left) and the helicity-angle distribution (right) in $\psi'\to J/\psi\pi^+\pi^-$. The BES, BESII and BESIII data sets are normalized using $\Gamma_{\psi'} = 293$~keV and ${\rm Br}(\psi' \to J/\psi \pi^{+}\pi^{-}) = 34.69\%$~\cite{ParticleDataGroup:2024cfk}.}
	\label{fit}
\end{figure*}

The same dispersive framework that is needed for the extraction of $\kappa$ and $|\alpha_{12}|$ also provides a natural description of the near-threshold $J/\psi{\cal P}$ amplitude. After crossing, the process $J/\psi{\cal P}\to J/\psi{\cal P}$ probes the $J/\psi J/\psi \to {\cal P}\bar{\cal P}$ amplitudes in the unphysical region around $s\le 0$. While the scattering length depends only on the threshold value of the amplitude, the effective range is controlled by its variation in the vicinity of threshold.

\begin{figure}[tb]
	\centering
	\includegraphics[width=0.48\textwidth]{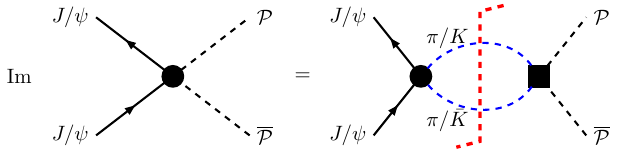}
	\caption{Unitarity relation for the amplitudes of the processes $J/\psi J/\psi \to {\cal P}\bar{\cal P}$. The black dots and squares denote the full $J/\psi J/\psi \to {\cal P}\bar{\cal P}$ amplitudes and the $\pi \pi$-$K\bar{K}$ coupled-channel rescattering, respectively. The red dashed line indicates the unitarity cut.}
	\label{fig:Unitarity}
\end{figure}

Following Refs.~\cite{Dong:2021lkh,Wu:2024xwy}, we include the $\pi\pi$-$K\bar{K}$ coupled-channel FSIs in the $J/\psi J/\psi \to {\cal P}\bar{\cal P}$ using the Muskhelishvili--Omn\`{e}s formalism~\cite{Omnes:1958hv,muskhelishvili2008singular}. 
The unitarity condition for the $S$-wave amplitudes can be written as~\cite{Donoghue:1990xh}
\begin{align}
	\operatorname{Im} {\bf M}_{0}(s) = \left[{\bf T}_{0}^{0}(s)\right]^{*}{\bm \Sigma}_{0}^{0}(s){\bf M}_{0}(s),
\end{align}
as illustrated diagrammatically in Fig.~\ref{fig:Unitarity}. Here, the column vector ${\bf M}_{0}(s) = \left( {\cal M}_{\pi,0}(s), {\cal M}_{K,0}(s) \right)^{\rm T}$ collects the $J/\psi J/\psi \to \bar{\cal P}{\cal P}$ amplitudes including FSIs. The diagonal phase-space matrix is
\begin{equation*}
    {\bm \Sigma}_{0}^{0}(s) \equiv \operatorname{diag}\left(\sigma_{\pi}(s)\theta(s-s_{\pi}), \sigma_{K}(s)\theta(s-s_{K})\right)
\end{equation*}
with $\sigma_{\cal P}(s) = \sqrt{1-4m_{\cal P}^{2}/s}$ and $s_{\cal P} = 4m_{\cal P}^{2}$.
The matrix ${\bf T}_{0}^{0}(s)$ denotes the coupled-channel $\pi\pi$-$K\bar{K}$ scattering amplitudes in the $IJ=00$ wave (for their definitions, see Refs.~\cite{Donoghue:1990xh, Hoferichter:2012wf}). 
The amplitudes are reconstructed via the dispersion relation
\begin{align}
	{\bf M}_{0}(s) = \frac{1}{\pi} \int_{4m_{\pi}^{2}}^{+\infty} \frac{\operatorname{Im} {\bf M}_{0}(s')}{s'-s-i\epsilon}\,{\rm d}s',
\end{align}
whose solution can be expressed in the Omn\`{e}s representation\footnote{Recently, BESIII reported a dip just above the $\pi\pi$ threshold in very-high-precision data~\cite{BESIII:2025ozb}, which can be well explained by including the $c_m$ term and a left-hand-cut contribution from $Z_c(3900)$ exchange~\cite{Chen:2025jip}. Since the $c_m$ and $Z_c(3900)$ contributions can be hardly distinguished~\cite{Chen:2025jip} and we do not aim for such precision, we neglect the $Z_c(3900)$ contribution.}
\begin{align}
	{\bf M}_{0}(s) = \bm{\Omega}_{0}^{0}(s){\bf P}_{0}(s).
\end{align}
The polynomial vector ${\bf P}_{0}(s)$ is determined by matching to the low-energy chiral amplitudes in Eq.~\eqref{S-amp}, while the $S$-wave Omn\`{e}s matrix $\bm{\Omega}_{0}^{0}$ is taken from Refs.~\cite{Hoferichter:2012wf, Cao:2024zlf}; its values for $s<0$ are obtained by analytical continuation using the same dispersion relation.

\begin{figure}[htb]
	\centering
	\includegraphics[width=0.45\textwidth]{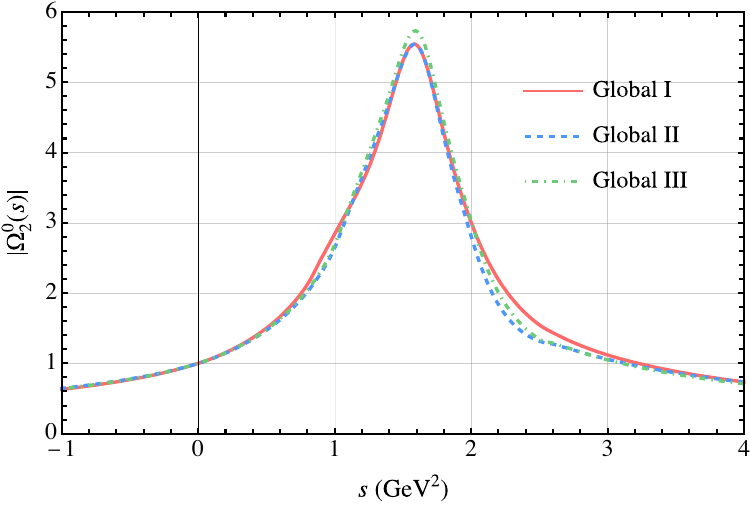}
	\caption{Modulus of the $D$-wave Omn\`es function $\Omega_{2}^{0}(s)$ as a function of $s$. The solid, dashed, and dot-dashed lines correspond to global fits I, II, and III of Ref.~\cite{Pelaez:2024uav}, respectively.}
	\label{OmnesD}
\end{figure}

For the $D$-wave, only the $\pi\pi$ channel is relevant in the region of interest. We write
\begin{align}
	{\cal M}_{\pi,2}(s) = \Omega_{2}^{0}(s)p_{\pi,2}(s),
\end{align}
where the polynomial $p_{\pi,2}(s)$ is fixed by matching to Eq.~\eqref{eq:D-amp}. Since the $f_{2}(1270)$ resonance dominates the isoscalar $D$-wave $\pi\pi$ scattering and induces a nonvanishing inelasticity, we construct the Omn\`{e}s function with the phase $\phi_{2}^{0}$ of the $\pi\pi$ partial wave $t_{2}^{0}$ rather than with the elastic phase shift $\delta_{2}^{0}$ alone~\cite{Hoferichter:2015hva},
\begin{align}
	\Omega_{2}^{0}(s) &= \exp \left[ \frac{s}{\pi} \int_{4m_{\pi}^{2}}^{\infty} \frac{{\rm d}s'}{s'} \frac{\phi_{2}^{0}(s')}{s'-s} \right],\\
	|t_{2}^{0}(s)|e^{i \phi_{2}^{0}(s)} &=\frac{\eta_{2}^{0}(s)e^{2i\delta_{2}^{0}(s)}-1}{2i \sigma_{\pi}(s)}. \label{eq:phi20}
\end{align}
Here $\delta_{2}^{0}$ and $\eta_{2}^{0}$ are taken from global fit~I of Ref.~\cite{Pelaez:2024uav} up to $s_{0}=(1.6~{\rm GeV})^{2}$ and are then smoothly extrapolated to infinity following the prescription of Ref.~\cite{Moussallam:1999aq}:
\begin{align}
	\delta_{2}^{0}(s) &= \pi + \left[\delta_{2}^{0}(s_{0})-\pi\right]f(s/s_{0}),\\
	\eta_{2}^{0}(s)   &=  1  + \left[  \eta_{2}^{0}(s_{0})- 1 \right]f(s/s_{0}),
\end{align}
with $f(x) = 2/(1+x^{3/2})$.
The resulting $\Omega_{2}^{0}(s)$ is shown in Fig.~\ref{OmnesD}.

For the kaon channel, the $D$-wave $K\bar K$ rescattering contribution is numerically suppressed near threshold, and we adopt a twice-subtracted dispersion relation that accounts for $\pi\pi$ rescatterings only\footnote{As a cross-check, we also perform the coupled-channel formalism using the $D$-wave Omn\`es matrix of Refs.~\cite{Cao:2025dkv,Danilkin:2025kyo}; the result is found to be compatible with those obtained from Eq.~\eqref{eq:2subDR_MK}.}
\begin{align}\label{eq:2subDR_MK}
    {\cal M}_{K,2}(s) = p_{K,2}(s) + \frac{s^{2}}{\pi}\int_{4m_{\pi}^{2}}^{\infty} \frac{\operatorname{Im} {\cal M}_{K,2}(s') }{s'^{2}(s'-s-i\epsilon)}\,{\rm d}s',
\end{align}
where the linear polynomial $p_{K,2}(s)$ is fixed by matching to Eq.~\eqref{eq:D-amp}. Here, the unitarity condition for the $D$-wave amplitudes implies that
\begin{align}
    \operatorname{Im} {\cal M}_{K,2}(s) = \left[g_{2}^{0}(s)\right]^{*} q_{\pi}^{4}(s) \sigma_{\pi}(s) \theta(s-s_{\pi}) {\cal M}_{\pi,2}(s),
\end{align}
with $q_{\pi}(s) = \sqrt{s/4 - m_{\pi}^{2}}$. Here, $g_{2}^{0}(s) = |g_{2}^{0}(s)| \exp[i \phi_{2}^{0}(s)]$ is the $\pi\pi \to K\bar{K}$ $D$-wave amplitude, where the modulus $|g_{2}^{0}(s)|$ is taken from the constrained-fit-to-data parameterization of Ref.~\cite{Pelaez:2020gnd}, and the phase $\phi_{2}^{0}(s)$ is taken to be the same as that in Eq.~\eqref{eq:phi20} under the single-channel approximation.

\begin{figure*}[tb]
	\centering
	\includegraphics[width=0.48\textwidth]{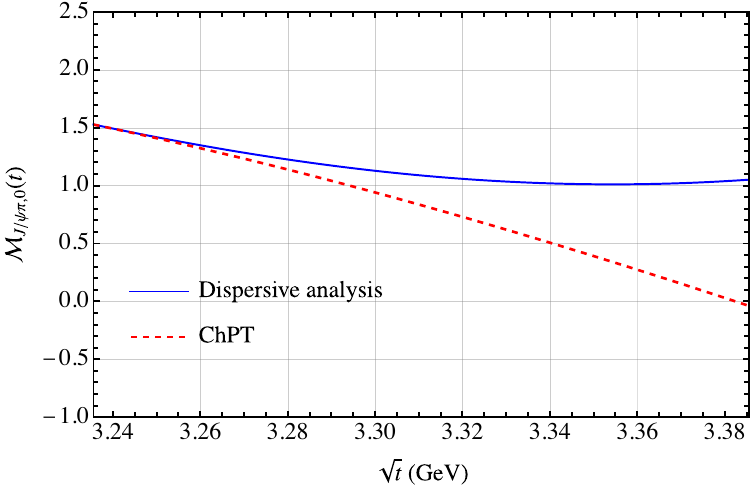}
	\includegraphics[width=0.48\textwidth]{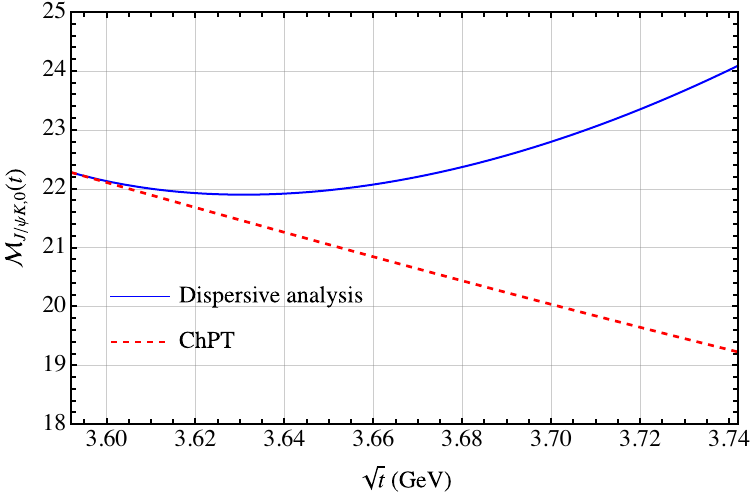}
	\caption{Near-threshold $S$-wave amplitudes for $J/\psi \pi \to J/\psi \pi$~(left) and $J/\psi K \to J/\psi K$~(right) from the dispersive representation~(blue solid) and from the tree-level chiral expression~(red dashed), shown for the reference choice $\alpha_{11}=|\alpha_{12}|$. In the present calculation, these amplitudes are real: the chiral input amplitudes are taken at tree level, and the Omn\`es functions are evaluated for $s(t,z_t)\leq 0$, below the crossed-channel physical cut.}
	\label{fig:MjpsiP}
\end{figure*}

The full crossed-channel amplitude then takes the form
\begin{align}\label{eq:MP-full}
	{\cal M}_{\cal P}(s,\cos\theta_s) = {\cal M}_{{\cal P},0}(s) + P_{2}(\cos\theta_s) {\cal M}_{{\cal P},2}(s).
\end{align}
Applying crossing symmetry, the $S$-wave $J/\psi\,{\cal P}$ amplitude can be written as\footnote{With the Mandelstam variables defined as $s=(p_{1}+p_{2})^{2}$, $t=(p_{1}-k_{1})^{2}$, and $u=(p_{1}-k_{2})^{2}$ for $J/\psi(p_{1})J/\psi(p_{2})\to \bar{\cal P}(k_{1}){\cal P}(k_{2})$, one has $s=0$ at $t=t_{\rm th}$ because $s+t+u=2M_{J/\psi}^{2}+2m_{\cal P}^{2}$ and $u_{\rm th}=(M_{J/\psi}-m_{\cal P})^{2}$.}
\begin{align}
	{\cal M}_{J/\psi{\cal P},0}(t) = \frac{1}{2}\int_{-1}^{1} {\cal M}_{\cal P}\left(s(t,z_t),\cos\theta_s(t,z_t) \right)\,{\rm d}z_t,
\end{align}
where $z_t\equiv \cos\theta_t$ and $\theta_t$ is the scattering angle in the $J/\psi{\cal P}$ center-of-mass frame:
\begin{align}
	s(t,z_t) &= -2q(t)^{2}(1-z_t),\label{eq:stz}\\
	\cos\theta_s(t,z_t) &= \frac{2t+s(t,z_t)-2M_{J/\psi}^{2}-2m_{\cal P}^{2}}{4\sqrt{s(t,z_t)/4-M_{J/\psi}^{2}}\sqrt{s(t,z_t)/4-m_{\cal P}^{2}}},\label{eq:zs}
\end{align}
where $q(t) = \lambda^{1/2}\left(t,M_{J/\psi}^{2},m_{\cal P}^{2}\right)/(2\sqrt{t})$ is the center-of-mass momentum of the $J/\psi {\cal P}$ system. This representation makes it straightforward to evaluate the amplitude slightly above threshold, where the effective-range expansion is applicable, as shown in Fig.~\ref{fig:MjpsiP}.

At the $J/\psi\,{\cal P}$ threshold, $t_{\rm th} = (M_{J/\psi}+m_{\cal P})^{2}$, the K\"{a}ll\'{e}n function vanishes, $\lambda(t_{\rm th},M_{J/\psi}^{2},m_{\cal P}^{2})=0$, implying that $s=0$ independently of the scattering angle according to Eq.~\eqref{eq:stz}.
With the analytic continuation from the physical $s$-channel region, the two square roots in Eq.~\eqref{eq:zs} are continued separately, so that $\cos\theta_s^{\rm th}=-1$ at $s=0$. 
Therefore the threshold projection becomes
\begin{align}
	{\cal M}^{\rm thr}_{J/\psi{\cal P}} = {\cal M}_{J/\psi{\cal P},0}(t_{\rm th}) = {\cal M}_{\cal P}(0,-1).
\end{align}
Here ${\cal M}_{\cal P}(0,-1)$ denotes the full crossed-channel
amplitude defined in Eq.~\eqref{eq:MP-full}, evaluated at this kinematic
point.
Because the subtraction polynomial is fixed by matching to the low-energy chiral amplitude, the dispersive representation reproduces the same threshold point, leading to the compact result
\begin{align}
	{\cal M}^{\rm thr}_{J/\psi{\cal P}}	= \frac{4\xi_{\cal P}m_{\cal P}^{2}}{F_{\pi}^{2}} \left(\tilde{c}_{1}^{(11)} + \tilde{c}_{2}^{(11)} - \tilde{c}_{m}^{(11)}\right).
	\label{eq:thresholdamp}
\end{align}
Note that this simple expression does not mean that the $\pi\pi$ interactions play no role here. The $\tilde c_i^{(11)}$ parameters are defined such that the $\pi\pi$ interaction Omn\`es matrices (function) equal to unity, and they should be regarded as the parameters renormalized by the $\pi\pi$ interactions, with the Omn\`es normalization corresponding to a particular regularization scheme.

To parameterize the threshold behavior, we relate the $S$-wave amplitude to the phase shift through
\begin{align}
	{\cal M}_{J/\psi{\cal P},0}(t)=\frac{8\pi\sqrt{t}}{q(t)\cot\delta_{0}(t)-iq(t)}.
\end{align}
The effective-range expansion then reads
\begin{align}
	q(t)\cot\delta_{0}(t) = -\frac{1}{a_{J/\psi{\cal P}}}+\frac{1}{2}r_{J/\psi{\cal P}}q^{2}(t)+{\cal O}(q^{4})\,.
\end{align}
Accordingly, the scattering length and effective range are obtained from
\begin{align}
	a_{J/\psi {\cal P}} &= -\frac{{\cal M}^{\rm thr}_{J/\psi{\cal P}}}{8\pi\left(M_{J/\psi} + m_{\cal P}\right)}\notag\\
	&=-\frac{\xi_{\cal P}\pi M_{J/\psi}m_{\cal P}^{2}}{2\beta_{0}\left(M_{J/\psi}+m_{\cal P}\right)}\left(2+9\kappa\right)\alpha_{11}\,,
	\label{eq:scaLen}\\
	r_{J/\psi {\cal P}} &= 16\pi \frac{t_{\rm th}}{M_{J/\psi}m_{\cal P}}\frac{\rm d}{{\rm d}t}\left[\sqrt{t}\operatorname{Re}\left(\frac{1}{{\cal M}_{J/\psi{\cal P},0}(t)}\right)\right]\bigg|_{t=t_{\rm th}}.\label{eq:effRan}
\end{align}
Equation~\eqref{eq:scaLen} shows explicitly that the scattering length depends only on the physical masses, $\kappa$, and the diagonal chromo-polarizability $\alpha_{11}$. The effective range, by contrast, probes the slope of the near-threshold amplitude and is therefore sensitive to the dispersive continuation embodied in ${\cal M}_{J/\psi{\cal P},0}(t)$.

Using the physical masses $m_{\pi} = 139.57~{\rm MeV}$, $m_{K}=496~{\rm MeV}$, and $M_{J/\psi}=3096~{\rm MeV}$, together with $\kappa=0.251$, $F_{\pi}=92.1~{\rm MeV}$, and the conservative constraint $\alpha_{11}\gtrsim|\alpha_{12}|=1.107~{\rm GeV}^{-3}$, we obtain the conservative upper-bound estimates
\begin{align}
	a_{J/\psi \pi} &\lesssim -0.0037~{\rm fm},& a_{J/\psi K} &\lesssim -0.049~{\rm fm}.
	\label{eq:gluon-res}
\end{align}
The same conservative input gives the corresponding effective-range estimates
\begin{align}
	r_{J/\psi \pi} &\lesssim 429~{\rm fm}\,,& r_{J/\psi K} &\lesssim 2.32~{\rm fm}\,. 
    \label{eq:ineq}
\end{align}
In our sign convention, a negative scattering length indicates that the $S$-wave $J/\psi\pi$ interaction is attractive while the attraction has not yet reached the critical strength needed to produce a bound state. The strong suppression of $a_{J/\psi\pi}$ reflects the combined effects of chiral and OZI suppressions, whereas the larger magnitude of $a_{J/\psi K}$ arises from the larger kaon mass and thus the stronger explicit chiral symmetry breaking associated with the strange quark mass. 
The seemingly large value of $r_{J/\psi\pi}$ mainly reflects the extremely small threshold amplitude, which enhances the derivative of $1/{\cal M}_{J/\psi\pi,0}(t)$ in Eq.~\eqref{eq:effRan}. Indeed, the relative size of the effective-range correction with respect to the leading scattering-length term scales as $|a_{J/\psi{\cal P}}r_{J/\psi{\cal P}}|q^2(t)/2$, so the very small value of $|a_{J/\psi\pi}|$ suppresses this correction near threshold. The large positive value of $r_{J/\psi\pi}$ is not in conflict with the Wigner bound; see, e.g., Ref.~\cite{Matuschek:2020gqe}. For an interaction with finite range $R$, the threshold form of the Wigner bound in our effective-range convention gives
\begin{align}
    r_{J/\psi{\cal P}}\leq 2R\left[1-\frac{R}{a_{J/\psi{\cal P}}}+\frac{R^2}{3a_{J/\psi{\cal P}}^2}\right].
\end{align}
Because $|a_{J/\psi\pi}|$ is extremely small, this upper bound becomes very weak due to the $R^3/a_{J/\psi\pi}^2$ term. Therefore, the numerically
large $r_{J/\psi\pi}$, if one takes a value towards the upper bound in Eq.~\eqref{eq:ineq}, should not be interpreted as evidence for a near-threshold pole.
By contrast, the kaon channel shows a more natural effective-range scale of hadronic size.

\subsection{Coupled-channel mechanism}

In the coupled-channel mechanism, the $J/\psi \pi$ and $J/\psi \bar{K}$ scattering processes proceed through intermediate coupling to open-charm channels. Specifically, the $J/\psi \pi$ system couples to the $D\bar{D}^*$ channel, while the $J/\psi \bar{K}$ system couples to the $D^*\bar{D}_s/D\bar{D}_s^*$ channels.\footnote{We do not distinguish between the $J/\psi K$ and $J/\psi \bar{K}$ scattering systems, since the corresponding potentials are identical.} This mechanism allows the scattering to occur via virtual transitions to these intermediate states, even when direct interactions are highly suppressed. The $S$-wave scattering amplitudes for these coupled-channel systems, encompassing both the $J/\psi {\cal P}$ and the corresponding open-charm $D\bar{D}^*$ ($D^*\bar{D}_s/D\bar{D}_s^*$) channels, have been constructed in Ref.~\cite{Du:2022jjv} to describe the exotic hadron candidates $Z_c(3900)$ and $Z_{cs}(3985)$, to which we refer the reader for further details. 
Since the direct $J/\psi{\cal P}\to J/\psi {\cal P}$ potential is not included in that work, the resulting scattering amplitudes effectively capture only the open-charm coupled-channel contributions, with the gluon-exchange mechanism excluded. 
For completeness, we briefly describe the framework for the $J/\psi\pi$-$D\bar{D}^*$ ($J/\psi \bar{K}$-$D^*\bar{D}_s/D\bar{D}_s^*$) scattering amplitude. Under SU(3) symmetry, the potential for them is constructed as 
\begin{align}
V_{(s)}(t)=\begin{pmatrix}
    0 & C_{12} \\
    C_{12} & C_Z+\frac{b}{2(m_{D_{(s)}}+m_{D^*})}\left[t-(m_{D_{(s)}}+m_{D^*})^2\right]
\end{pmatrix},
\end{align} 
where $C_Z$, $C_{12}$, and $b$ are low-energy constants, the subscript $(s)$ denotes the $J/\psi\pi$-$D\bar{D}^*$ ($J/\psi \bar{K}$-$D^*\bar{D}_s/D\bar{D}_s^*$) system. When the SU(3) light flavor symmetry breaking effects are considered, the $C_Z$ and $C_{12}$ for the strange system ($J/\psi \bar{K}$-$D^*\bar{D}_s/D\bar{D}_s^*$) are replaced by $C_Z^s=C_Z+\delta Z$ and $C_{12}^s=C_{12}+\delta_{12}$. Apparently, the SU(3) symmetry corresponds to the special case $\delta Z=\delta_{12}=0$. 

The coupled-channel $T$-matrix is obtained by\footnote{The sign convention for the $T$-matrix in this article differs by an overall minus sign from that of Ref.~\cite{Du:2022jjv}.} 
\begin{align}
    T_{(s)}(t)=-\left[1-V_{(s)}(t)\cdot G_{(s)}(t)\right]^{-1}\cdot V_{(s)}(t)\, ,
\end{align}
where the $G_{(s)}$ is the loop-function diagonal matrix $G(t)={\rm diag}\left\{G_1(t),G_2(t)\right\}$, 
\begin{align}
    G_i(t) &= i\int\frac{{\rm d}^4q}{(2\pi)^4}\frac{1}{(q^2-m_{i,1}^2)\left((p-q)^2-m_{i,2}^2\right)}\,,&p^2&=t\,,\notag
\end{align}
with $m_{i,1}$ and $m_{i,2}$ the masses of the two intermediate states in the $i$th channel. The loop function $G_i(t)$ is evaluated with a once-subtracted dispersion relation, and its explicit expression reads
\begin{align}
\operatorname{Re}G_i(t)=&\,\frac{1}{16\pi^2}\Bigg[ a_i(\mu) + \log \frac{m_{i,1}^2}{\mu^2}+\frac{t-m_{i,1}^2+m_{i,2}^2}{2t}\log\frac{m_{i,2}^2}{m_{i,1}^2} \notag\\
&\,+\frac{\sigma_i(t)}{2t}\log \frac{t+\sigma_i(t)-m_{i,1}^2-m_{i,2}^2}{t-\sigma_i(t)-m_{i,1}^2-m_{i,2}^2}\Bigg]\,,\\
\operatorname{Im}G_i(t) =&\,-\frac{\sigma_i(t)}{16\pi t}\,,
\end{align}
with $\sigma_i(t)=\lambda^{1/2}(t,m_{i,1}^{2},m_{i,2}^{2})$ and $\mu=1~{\rm GeV}$ the renormalization scale. The $S$-wave scattering lengths for the $J/\psi \pi$ and $J/\psi \bar K$ are related to the scattering amplitudes at their threshold
\begin{align}
T_{(s),11}(t=t_{\rm th}) =  -8\pi(M_{J/\psi}+m_{\cal P}) a_{J/\psi{\cal P} }^{\rm CC}\,.
\end{align}

Adopting the parameters determined in Ref.~\cite{Du:2022jjv} by fitting the BESIII data on $e^+e^-\to J/\psi \pi^+\pi^-$~\cite{BESIII:2017bua}, $e^+e^-\to \pi^+(D^0D^{*-})$~\cite{BESIII:2015pqw}, and $e^+e^-\to K^+(D^{*0}D_s^-+D^0D_s^{*-})$~\cite{BESIII:2020qkh}, we obtain the $S$-wave $J/\psi{\cal P}$ scattering lengths from the coupled-channel mechanism as
\begin{align}
    a_{J/\psi\pi}^{\rm CC} &\in [-4.2, 6.1]\times 10^{-6} ~{\rm fm}\,, \notag\\
    a_{J/\psi K}^{\rm CC}  &\in  [-0.01, -8.9\times 10^{-6}] ~{\rm fm}\,.
	\label{eq:cc-res}
\end{align}
The sizable uncertainty in $a_{J/\psi K}^\text{CC}$ originates from significant SU(3) flavor-symmetry breaking in the parameter $C_{12}$ within scheme~(c) of Sec.~IV in Ref.~\cite{Du:2022jjv}, where $\delta_{12}\gg C_{12}$. Since $C_{12}$ governs the transition potential between the $J/\psi {\cal P}$ and open-charm channels, its value strongly influences the resulting scattering length. 

Comparing the results in Eqs.~\eqref{eq:gluon-res} and \eqref{eq:cc-res}, one finds that the soft-gluon exchange mechanism provides the dominant contribution to $J/\psi {\cal P}$ scattering in the near-threshold region, which is analogous to the finding in Ref.~\cite{Wu:2024xwy} for $J/\psi$-nucleon scattering.

\section{Summary}
\label{sec4}

In this work, we have investigated the low-energy scattering of $\pi$ and $K$ mesons off $J/\psi$. We demonstrate that the symmetry-breaking terms in the charmonium--pNG-boson interaction Lagrangian induce mixing between the bare charmonium fields, rendering them non-diagonal in the mass basis. Upon diagonalization, the physical $J/\psi$ and $\psi'$ states are identified, and the symmetry-breaking terms remain nonvanishing for terms with at least two pNG fields, contributing to the transition amplitude of $\psi'\to J/\psi\pi\pi$.

Using the resulting effective Lagrangian, we construct the crossed-channel amplitudes for $J/\psi J/\psi \to {\cal P}\bar{\cal P}$ and incorporated the correlated $\pi\pi$ and $K\bar K$ rescattering effects through the Muskhelishvili--Omn\`{e}s formalism. The same dispersive framework is used to extract $\kappa$ and $|\alpha_{12}|$ from $\psi'\to J/\psi\pi\pi$ and to continue the crossed amplitudes to the near-threshold $J/\psi{\cal P}$ region. 

With the conservative bound $\alpha_{11}\gtrsim |\alpha_{12}|$ and the updated central values of $\kappa$ and $|\alpha_{12}|$ extracted from extracted from $\psi' \to J/\psi\pi\pi$, we obtain the upper-bound estimates $a_{J/\psi\pi}\lesssim -0.0037$~fm and $a_{J/\psi K}\lesssim -0.049$~fm. The same conservative input gives the corresponding effective ranges estimates $r_{J/\psi\pi}\lesssim 429$~fm and $r_{J/\psi K}\lesssim 2.32$~fm. The $J/\psi\pi$ interaction is therefore extremely weak near threshold, consistent with the combined chiral and OZI suppressions, while the larger magnitude of $a_{J/\psi K}$ reflects the stronger explicit chiral symmetry breaking associated with the strange quark mass. In contrast, the contributions from the coupled-channel mechanism, where $J/\psi\pi$ and  $J/\psi \bar{K}$ couple to open-charmed channels, are found to be significantly smaller than those from soft-gluon exchange, indicating that the multi-gluon exchange mechanism dominates the low-energy $J/\psi\pi$ and $J/\psi K$ scattering. These predictions can be tested against forthcoming lattice QCD calculations.

Combined with the analogous conclusion reached in Ref.~\cite{Wu:2024xwy} for $J/\psi$-nucleon scattering, the dominance of the multi-gluon exchange mechanism over the coupled-channel mechanism may be a universal feature for the scattering of light hadrons off charmonia. Whether this conclusion extends to bottomonia remains an open question, since the chromo-polarizabilities of fully-bottom systems are expected to be considerably smaller than those of fully-charm systems~\cite{Dong:2022rwr}.

\section*{Acknowledgements} 

This work is supported in part by the National Natural Science Foundation of China under Grants No. 12547115, No. 12125507, No. 12361141819, No. 12447101, and No. 12547111; by the National Key R\&D Program of China under Grant No. 2023YFA1606703; and by the Chinese Academy of Sciences under Grant No. YSBR-101. MLD gratefully acknowledges the support of the Peng Huan-Wu Visiting Professorship and the hospitality of the Institute of Theoretical Physics, Chinese Academy of Sciences, where part of this work was completed.

\bibliographystyle{elsarticle-num} 
\bibliography{refs.bib}

@article{Barnes:2003vt,
    author = "Barnes, T.",
    editor = "Elster, C. and Speth, J. and Walcher, T.",
    title = "{Charmonium cross-sections and the QGP}",
    eprint = "nucl-th/0306031",
    archivePrefix = "arXiv",
    doi = "10.1140/epja/i2002-10276-4",
    journal = "Eur. Phys. J. A",
    volume = "18",
    pages = "531",
    year = "2003"
}

@article{Peskin:1979va,
    author = "Peskin, Michael E.",
    title = "{Short Distance Analysis for Heavy Quark Systems. 1. Diagrammatics}",
    reportNumber = "HUTP-79/A008",
    doi = "10.1016/0550-3213(79)90199-8",
    journal = "Nucl. Phys. B",
    volume = "156",
    pages = "365--390",
    year = "1979"
}

@article{Bhanot:1979vb,
    author = "Bhanot, Gyan and Peskin, Michael E.",
    title = "{Short Distance Analysis for Heavy Quark Systems. 2. Applications}",
    reportNumber = "HUTP-79/A015",
    doi = "10.1016/0550-3213(79)90200-1",
    journal = "Nucl. Phys. B",
    volume = "156",
    pages = "391--416",
    year = "1979"
}

@article{Okubo:1963fa,
    author = "Okubo, S.",
    title = "{$\varphi$-meson and unitary symmetry model}",
    doi = "10.1016/S0375-9601(63)92548-9",
    journal = "Phys. Lett.",
    volume = "5",
    pages = "165--168",
    year = "1963"
}

@article{Zweig:1964jf,
    author = "Zweig, G.",
    title = "{An SU(3) model for strong interaction symmetry and its breaking. Version 2}",
    journal = "{Developments in the Quark Theory of Hadrons}",
    volume = "1",
    reportNumber = "CERN-TH-412",
    doi = "10.17181/CERN-TH-412",
    pages = "22--101",
    month = "2",
    year = "1964"
}

@article{Iizuka:1966fk,
    author = "Iizuka, Jugoro",
    title = "{Systematics and phenomenology of meson family}",
    doi = "10.1143/PTPS.37.21",
    journal = "Prog. Theor. Phys. Suppl.",
    volume = "37",
    pages = "21--34",
    year = "1966"
}

@article{Brodsky:1989jd,
    author = "Brodsky, Stanley J. and Schmidt, I. A. and de Teramond, G. F.",
    title = "{Nuclear Bound Quarkonium}",
    reportNumber = "SLAC-PUB-5102",
    doi = "10.1103/PhysRevLett.64.1011",
    journal = "Phys. Rev. Lett.",
    volume = "64",
    pages = "1011",
    year = "1990"
}

@article{Brodsky:1997gh,
    author = "Brodsky, Stanley J. and Miller, Gerald A.",
    title = "{Is $J/\psi$-nucleon scattering dominated by the gluonic van der Waals interaction?}",
    eprint = "hep-ph/9707382",
    archivePrefix = "arXiv",
    reportNumber = "SLAC-PUB-7553, DOE-ER-40561-330, INT97-19-07, DOE-ER-41014-16-N97",
    doi = "10.1016/S0370-2693(97)01045-9",
    journal = "Phys. Lett. B",
    volume = "412",
    pages = "125--130",
    year = "1997"
}

@article{Gottfried:1977gp,
    author = "Gottfried, Kurt",
    title = "{Hadronic Transitions Between Quark-Antiquark Bound States}",
    reportNumber = "CLNS-381",
    doi = "10.1103/PhysRevLett.40.598",
    journal = "Phys. Rev. Lett.",
    volume = "40",
    pages = "598",
    year = "1978"
}

@article{Voloshin:1978hc,
    author = "Voloshin, M. B.",
    title = "{On Dynamics of Heavy Quarks in Nonperturbative QCD Vacuum}",
    reportNumber = "ITEP-86-1978",
    doi = "10.1016/0550-3213(79)90037-3",
    journal = "Nucl. Phys. B",
    volume = "154",
    pages = "365--380",
    year = "1979"
}

@article{Mannel:1995jt,
    author = "Mannel, Thomas and Urech, Res",
    title = "{Hadronic decays of excited heavy quarkonia}",
    eprint = "hep-ph/9510406",
    archivePrefix = "arXiv",
    reportNumber = "TTP-95-36",
    doi = "10.1007/s002880050344",
    journal = "Z. Phys. C",
    volume = "73",
    pages = "541--546",
    year = "1997"
}

@article{Brown:1975dz,
  title = {Chiral Symmetry and $\psi' \to \psi\pi\pi$ Decay},
  author = {Brown, Lowell S. and Cahn, Robert N.},
  year = 1975,
  journal = {Physical Review Letters},
  volume = {35},
  pages = {1},
  doi = {10.1103/PhysRevLett.35.1}
}

@article{Luke:1992tm,
    author = "Luke, Michael E. and Manohar, Aneesh V. and Savage, Martin J.",
    title = "{A QCD Calculation of the interaction of quarkonium with nuclei}",
    eprint = "hep-ph/9204219",
    archivePrefix = "arXiv",
    reportNumber = "UCSD-PTH-92-12",
    doi = "10.1016/0370-2693(92)91114-O",
    journal = "Phys. Lett. B",
    volume = "288",
    pages = "355--359",
    year = "1992"
}

@article{Sibirtsev:2005ex,
    author = "Sibirtsev, A. and Voloshin, M. B.",
    title = "{The Interaction of slow $J/\psi$ and $\psi'$ with nucleons}",
    eprint = "hep-ph/0502068",
    archivePrefix = "arXiv",
    reportNumber = "FZJ-IKP-TH-2005-6, FTPI-MINN-05-03, UMN-TH-2344-05",
    doi = "10.1103/PhysRevD.71.076005",
    journal = "Phys. Rev. D",
    volume = "71",
    pages = "076005",
    year = "2005"
}

@article{Kaidalov:1992hd,
    author = "Kaidalov, A. B. and Volkovitsky, P. E.",
    title = "{Heavy quarkonia interactions with nucleons and nuclei}",
    doi = "10.1103/PhysRevLett.69.3155",
    journal = "Phys. Rev. Lett.",
    volume = "69",
    pages = "3155--3156",
    year = "1992"
}

@article{deTeramond:1997ny,
    author = "de Teramond, Guy F. and Espinoza, Randall and Ortega-Rodriguez, Manuel",
    title = "{Proton proton spin correlations at charm threshold and quarkonium bound to nuclei}",
    eprint = "hep-ph/9708202",
    archivePrefix = "arXiv",
    doi = "10.1103/PhysRevD.58.034012",
    journal = "Phys. Rev. D",
    volume = "58",
    pages = "034012",
    year = "1998"
}

@article{Beane:2014sda,
    author = "Beane, S. R. and Chang, E. and Cohen, S. D. and Detmold, W. and Lin, H. -W. and Orginos, K. and Parre{\~n}o, A. and Savage, M. J.",
    title = "{Quarkonium-Nucleus Bound States from Lattice QCD}",
    eprint = "1410.7069",
    archivePrefix = "arXiv",
    primaryClass = "hep-lat",
    reportNumber = "INT-PUB-14-049, NT@UW-14-23, MIT-CTP-4600, JLAB-THY-14-1992",
    doi = "10.1103/PhysRevD.91.114503",
    journal = "Phys. Rev. D",
    volume = "91",
    number = "11",
    pages = "114503",
    year = "2015"
}

@article{Krein:2020yor,
    author = "Krein, G. and Peixoto, T. C.",
    title = "{Femtoscopy of the Origin of the Nucleon Mass}",
    eprint = "2011.11615",
    archivePrefix = "arXiv",
    primaryClass = "hep-ph",
    doi = "10.1007/s00601-020-01581-1",
    journal = "Few Body Syst.",
    volume = "61",
    number = "4",
    pages = "49",
    year = "2020"
}

@article{Chen:1997zza,
    author = "Chen, Jiunn-Wei and Savage, Martin J.",
    title = "{Hadronic and electromagnetic interactions of quarkonia}",
    eprint = "hep-ph/9710338",
    archivePrefix = "arXiv",
    reportNumber = "DOE-ER-40427-31-N96",
    doi = "10.1103/PhysRevD.57.2837",
    journal = "Phys. Rev. D",
    volume = "57",
    pages = "2837--2846",
    year = "1998"
}

@article{Yokokawa:2006td,
    author = "Yokokawa, Kazuo and Sasaki, Shoichi and Hatsuda, Tetsuo and Hayashigaki, Arata",
    title = "{First lattice study of low-energy charmonium-hadron interaction}",
    eprint = "hep-lat/0605009",
    archivePrefix = "arXiv",
    doi = "10.1103/PhysRevD.74.034504",
    journal = "Phys. Rev. D",
    volume = "74",
    pages = "034504",
    year = "2006"
}

@article{Liu:2008rza,
    author = "Liu, Liuming and Lin, Huey-Wen and Orginos, Kostas",
    editor = "Aubin, Christopher and Cohen, Saul and Dawson, Chris and Dudek, Jozef and Edwards, Robert and Joo, Balint and Lin, Huey-Wen and Orginos, Kostas and Richards, David and Thacker, Hank",
    title = "{Charmed Hadron Interactions}",
    eprint = "0810.5412",
    archivePrefix = "arXiv",
    primaryClass = "hep-lat",
    reportNumber = "JLAB-THY-08-899",
    doi = "10.22323/1.066.0112",
    journal = "PoS",
    volume = "LATTICE2008",
    pages = "112",
    year = "2008"
}

@article{Liu:2012dv,
    author = "Liu, Xiao-Hai and Guo, Feng-Kun and Epelbaum, Evgeny",
    title = "{Extracting $\pi\pi$ $S$-wave scattering lengths from cusp effect in heavy quarkonium dipion transitions}",
    eprint = "1212.4066",
    archivePrefix = "arXiv",
    primaryClass = "hep-ph",
    doi = "10.1140/epjc/s10052-013-2284-2",
    journal = "Eur. Phys. J. C",
    volume = "73",
    number = "1",
    pages = "2284",
    year = "2013"
}

@article{Chen:2015jgl,
    author = "Chen, Yun-Hua and Daub, Johanna T. and Guo, Feng-Kun and Kubis, Bastian and Mei{\ss}ner, Ulf-G. and Zou, Bing-Song",
    title = "{Effect of $Z_b$ states on $\Upsilon(3S)\to\Upsilon(1S)\pi\pi$ decays}",
    eprint = "1512.03583",
    archivePrefix = "arXiv",
    primaryClass = "hep-ph",
    doi = "10.1103/PhysRevD.93.034030",
    journal = "Phys. Rev. D",
    volume = "93",
    number = "3",
    pages = "034030",
    year = "2016"
}

@article{Chen:2019hmz,
    author = "Chen, Yun-Hua",
    title = "{Chromopolarizability of Charmonium and $\pi\pi$ Final State Interaction Revisited}",
    eprint = "1901.04126",
    archivePrefix = "arXiv",
    primaryClass = "hep-ph",
    doi = "10.1155/2019/7650678",
    journal = "Adv. High Energy Phys.",
    volume = "2019",
    pages = "7650678",
    year = "2019"
}

@article{Dong:2021lkh,
    author = "Dong, Xiang-Kun and Baru, Vadim and Guo, Feng-Kun and Hanhart, Christoph and Nefediev, Alexey and Zou, Bing-Song",
    title = "{Is the existence of a $J/\psi J/\psi$ bound state plausible?}",
    eprint = "2107.03946",
    archivePrefix = "arXiv",
    primaryClass = "hep-ph",
    doi = "10.1016/j.scib.2021.09.009",
    journal = "Sci. Bull.",
    volume = "66",
    number = "24",
    pages = "2462--2470",
    year = "2021"
}

@article{Wu:2024xwy,
    author = "Wu, Bing and Dong, Xiang-Kun and Du, Meng-Lin and Guo, Feng-Kun and Zou, Bing-Song",
    title = "{Deciphering the mechanism of $J/\psi$-nucleon scattering}",
    eprint = "2410.19526",
    archivePrefix = "arXiv",
    primaryClass = "hep-ph",
    doi = "10.1016/j.fmre.2025.07.005",
    journal = "Fund. Res.",
    volume = "5",
    pages = "2530--2536",
    year = "2025"
}

@article{Brambilla:2015rqa,
    author = "Brambilla, Nora and Krein, Gast{\~a}o and Tarr{\'u}s Castell{\`a}, Jaume and Vairo, Antonio",
    title = "{Long-range properties of $1S$ bottomonium states}",
    eprint = "1510.05895",
    archivePrefix = "arXiv",
    primaryClass = "hep-ph",
    reportNumber = "TUM-EFT-70-15",
    doi = "10.1103/PhysRevD.93.054002",
    journal = "Phys. Rev. D",
    volume = "93",
    number = "5",
    pages = "054002",
    year = "2016"
}

@article{Voloshin:2007dx,
    author = "Voloshin, M. B.",
    title = "{Charmonium}",
    eprint = "0711.4556",
    archivePrefix = "arXiv",
    primaryClass = "hep-ph",
    reportNumber = "FTPI-MINN-07-34, UMN-TH-2625-07",
    doi = "10.1016/j.ppnp.2008.02.001",
    journal = "Prog. Part. Nucl. Phys.",
    volume = "61",
    pages = "455--511",
    year = "2008"
}

@article{Novikov:1980fa,
    author = "Novikov, V. A. and Shifman, Mikhail A.",
    title = "{Comment on the $\psi' \to J/\psi \pi \pi$  Decay}",
    reportNumber = "ITEP-93-1980",
    doi = "10.1007/BF01429829",
    journal = "Z. Phys. C",
    volume = "8",
    pages = "43",
    year = "1981"
}

@article{Eides:2015dtr,
    author = "Eides, Michael I. and Petrov, Victor Yu. and Polyakov, Maxim V.",
    title = "{Narrow Nucleon-$\psi(2S)$ Bound State and LHCb Pentaquarks}",
    eprint = "1512.00426",
    archivePrefix = "arXiv",
    primaryClass = "hep-ph",
    doi = "10.1103/PhysRevD.93.054039",
    journal = "Phys. Rev. D",
    volume = "93",
    number = "5",
    pages = "054039",
    year = "2016"
}

@article{Eides:2017xnt,
    author = "Eides, Michael I. and Petrov, Victor Yu. and Polyakov, Maxim V.",
    title = "{Pentaquarks with hidden charm as hadroquarkonia}",
    eprint = "1709.09523",
    archivePrefix = "arXiv",
    primaryClass = "hep-ph",
    doi = "10.1140/epjc/s10052-018-5530-9",
    journal = "Eur. Phys. J. C",
    volume = "78",
    number = "1",
    pages = "36",
    year = "2018"
}

@article{TarrusCastella:2018php,
    author = "Tarr{\'u}s Castell{\`a}, Jaume and Krein, Gast{\~a}o",
    title = "{Effective field theory for the nucleon-quarkonium interaction}",
    eprint = "1803.05412",
    archivePrefix = "arXiv",
    primaryClass = "hep-ph",
    reportNumber = "TUM-EFT 108/18, TUM-EFT-108-18",
    doi = "10.1103/PhysRevD.98.014029",
    journal = "Phys. Rev. D",
    volume = "98",
    number = "1",
    pages = "014029",
    year = "2018"
}

@article{Kawanai:2010ev,
    author = "Kawanai, Taichi and Sasaki, Shoichi",
    title = "{Charmonium-nucleon potential from lattice QCD}",
    eprint = "1009.3332",
    archivePrefix = "arXiv",
    primaryClass = "hep-lat",
    reportNumber = "TKYNT-10-14",
    doi = "10.1103/PhysRevD.82.091501",
    journal = "Phys. Rev. D",
    volume = "82",
    pages = "091501",
    year = "2010"
}

@article{Sugiura:2017vks,
    author = "Sugiura, Takuya and Ikeda, Yoichi and Ishii, Noriyoshi",
    editor = "Della Morte, M. and Fritzsch, P. and G{\'a}miz S{\'a}nchez, E. and Pena Ruano, C.",
    title = "{Charmonium-nucleon interactions from the time-dependent HAL QCD method}",
    eprint = "1711.11219",
    archivePrefix = "arXiv",
    primaryClass = "hep-lat",
    doi = "10.1051/epjconf/201817505011",
    journal = "EPJ Web Conf.",
    volume = "175",
    pages = "05011",
    year = "2018"
}

@article{Polyakov:2018aey,
    author = "Polyakov, Maxim V. and Schweitzer, Peter",
    title = "{Determination of $J/\psi$ chromoelectric polarizability from lattice data}",
    eprint = "1801.08984",
    archivePrefix = "arXiv",
    primaryClass = "hep-ph",
    doi = "10.1103/PhysRevD.98.034030",
    journal = "Phys. Rev. D",
    volume = "98",
    number = "3",
    pages = "034030",
    year = "2018"
}

@article{Voloshin:1979uv,
    author = "Voloshin, M. B.",
    title = "{Precoulombic Asymptotics for Energy Levels of Heavy Quarkonium}",
    reportNumber = "ITEP-54-1979",
    journal = "Sov. J. Nucl. Phys.",
    volume = "36",
    pages = "143",
    year = "1982"
}

@article{BES:1999guu,
    author = "Bai, J. Z. and others",
    collaboration = "BES",
    title = "{$\psi(2S) \to \pi^{+} \pi^{-} J / \psi$ decay distributions}",
    eprint = "hep-ex/9909038",
    archivePrefix = "arXiv",
    reportNumber = "SLAC-REPRINT-1999-088, BIHEP-EP1-99-03, UH511-939-99",
    doi = "10.1103/PhysRevD.62.032002",
    journal = "Phys. Rev. D",
    volume = "62",
    pages = "032002",
    year = "2000"
}

@article{BES:2006eer,
    author = "Ablikim, M. and others",
    collaboration = "BES",
    title = "{Production of sigma in $\psi(2S) \to \pi^{+} \pi^{-} J/\psi$}",
    eprint = "hep-ex/0610023",
    archivePrefix = "arXiv",
    doi = "10.1016/j.physletb.2006.11.056",
    journal = "Phys. Lett. B",
    volume = "645",
    pages = "19--25",
    year = "2007"
}

@article{ATLAS:2016kwu,
    author = "Aaboud, Morad and others",
    collaboration = "ATLAS",
    title = "{Measurements of $\psi(2S)$ and $X(3872) \to J/\psi\pi^+\pi^-$ production in $pp$ collisions at $\sqrt{s} = 8$ TeV with the ATLAS detector}",
    eprint = "1610.09303",
    archivePrefix = "arXiv",
    primaryClass = "hep-ex",
    reportNumber = "CERN-EP-2016-193",
    doi = "10.1007/JHEP01(2017)117",
    journal = "JHEP",
    volume = "01",
    pages = "117",
    year = "2017"
}

@article{Donoghue:1990xh,
    author = "Donoghue, John F. and Gasser, J. and Leutwyler, H.",
    title = "{The Decay of a Light Higgs Boson}",
    reportNumber = "CERN-TH-5644/90, BUTP-89-06",
    doi = "10.1016/0550-3213(90)90474-R",
    journal = "Nucl. Phys. B",
    volume = "343",
    pages = "341--368",
    year = "1990"
}

@article{Hoferichter:2012wf,
    author = "Hoferichter, M. and Ditsche, C. and Kubis, B. and Mei{\ss}ner, U. G.",
    title = "{Dispersive analysis of the scalar form factor of the nucleon}",
    eprint = "1204.6251",
    archivePrefix = "arXiv",
    primaryClass = "hep-ph",
    doi = "10.1007/JHEP06(2012)063",
    journal = "JHEP",
    volume = "06",
    pages = "063",
    year = "2012"
}

@article{Omnes:1958hv,
    author = "Omn\`es, R.",
    title = "{On the Solution of certain singular integral equations of quantum field theory}",
    doi = "10.1007/BF02747746",
    journal = "Nuovo Cim.",
    volume = "8",
    pages = "316--326",
    year = "1958"
}

@book{muskhelishvili2008singular,
  title = "{Singular Integral Equations: Boundary Problems of Function Theory and Their Application to Mathematical Physics}",
  author = "{Muskhelishvili, N. I.}",
  year = "1958",
  edition = {1},
  publisher = {Springer Dordrecht},
  doi = "10.1007/978-94-009-9994-7"
}

@article{Cao:2024zlf,
    author = "Cao, Xiong-Hui and Guo, Feng-Kun and Li, Qu-Zhi and Yao, De-Liang",
    title = "{Dispersive determination of nucleon gravitational form factors}",
    eprint = "2411.13398",
    archivePrefix = "arXiv",
    primaryClass = "hep-ph",
    doi = "10.1038/s41467-025-62278-9",
    journal = "Nature Commun.",
    volume = "16",
    pages = "6979",
    year = "2025"
}

@article{Pineda:2019mhw,
	author = "Pineda, Antonio and Tarr{\'u}s Castell{\`a}, Jaume",
	title = "{Novel implementation of the multipole expansion to quarkonium hadronic transitions}",
	eprint = "1905.03794",
	archivePrefix = "arXiv",
	primaryClass = "hep-ph",
	doi = "10.1103/PhysRevD.100.054021",
	journal = "Phys. Rev. D",
	volume = "100",
	number = "5",
	pages = "054021",
	year = "2019"
}

@article{Du:2022jjv,
    author = "Du, Meng-Lin and Albaladejo, Miguel and Guo, Feng-Kun and Nieves, Juan",
    title = "{Combined analysis of the $Z_c(3900)$ and the $Z_{cs}(3985)$ exotic states}",
    eprint = "2201.08253",
    archivePrefix = "arXiv",
    primaryClass = "hep-ph",
    doi = "10.1103/PhysRevD.105.074018",
    journal = "Phys. Rev. D",
    volume = "105",
    number = "7",
    pages = "074018",
    year = "2022"
}

@article{BESIII:2017bua,
    author = "Ablikim, Medina and others",
    collaboration = "BESIII",
    title = "{Determination of the Spin and Parity of the $Z_c(3900)$}",
    eprint = "1706.04100",
    archivePrefix = "arXiv",
    primaryClass = "hep-ex",
    doi = "10.1103/PhysRevLett.119.072001",
    journal = "Phys. Rev. Lett.",
    volume = "119",
    number = "7",
    pages = "072001",
    year = "2017"
}

@article{BESIII:2015pqw,
    author = "Ablikim, M. and others",
    collaboration = "BESIII",
    title = "{Confirmation of a charged charmoniumlike state $Z_c(3885)^{\mp}$ in $e^+e^-\to\pi^{\pm}(D\bar{D}^*)^\mp$ with double $D$ tag}",
    eprint = "1509.01398",
    archivePrefix = "arXiv",
    primaryClass = "hep-ex",
    doi = "10.1103/PhysRevD.92.092006",
    journal = "Phys. Rev. D",
    volume = "92",
    number = "9",
    pages = "092006",
    year = "2015"
}

@article{BESIII:2020qkh,
    author = "Ablikim, Medina and others",
    collaboration = "BESIII",
    title = "{Observation of a Near-Threshold Structure in the $K^+$ Recoil-Mass Spectra in $e^+e^- \to K^+(D_s^-D^{*0}+D_s^{*-}D^0$)}",
    eprint = "2011.07855",
    archivePrefix = "arXiv",
    primaryClass = "hep-ex",
    doi = "10.1103/PhysRevLett.126.102001",
    journal = "Phys. Rev. Lett.",
    volume = "126",
    number = "10",
    pages = "102001",
    year = "2021"
}

@article{BESIII:2025ozb,
    author = "Ablikim, Medina and others",
    collaboration = "BESIII",
    title = "{Observation of a Threshold Enhancement in the $\pi^{+}\pi^-$ Spectrum in $\psi(3686)\to \pi^{+}\pi^{-}{J/\psi}$ Decays}",
    eprint = "2509.23761",
    archivePrefix = "arXiv",
    primaryClass = "hep-ex",
    doi = "10.1103/5h6z-p6nl",
    journal = "Phys. Rev. Lett.",
    volume = "136",
    number = "14",
    pages = "141902",
    year = "2026"
}

@article{Chen:2025jip,
    author = "Chen, Yun-Hua and Dong, Xiang-Kun and Guo, Feng-Kun and Hanhart, Christoph and Kubis, Bastian",
    title = "{Decoding the structure near the $\pi^+\pi^-$ mass threshold in $\psi(3686) \to J/\psi \pi^+ \pi^-$ decays}",
    eprint = "2512.01679",
    archivePrefix = "arXiv",
    primaryClass = "hep-ph",
    doi = "10.1103/r5lz-864p",
    journal = "Phys. Rev. D",
    volume = "113",
    number = "5",
    pages = "054018",
    year = "2026"
}

@article{Guo:2006ya,
  title = {Chromo-polarizability and {$\pi\pi$} final state interaction},
  author = {Guo, Feng-Kun and Shen, Peng-Nian and Chiang, Huan-Ching},
  year = 2006,
  journal = {Phys. Rev. D},
  volume = {74},
  eprint = {hep-ph/0604252},
  pages = {014011},
  doi = {10.1103/PhysRevD.74.014011},
  archiveprefix = {arXiv}
}

@article{Dong:2022rwr,
  title = {Chromopolarizabilities of fully heavy baryons},
  author = {Dong, Xiang-Kun and Guo, Feng-Kun and Nefediev, Alexey and Castell{\`a}, Jaume Tarr{\'u}s},
  year = 2023,
  journal = {Phys. Rev. D},
  volume = {107},
  eprint = {2211.14100},
  primaryclass = {hep-ph},
  pages = {034020},
  doi = {10.1103/PhysRevD.107.034020},
  archiveprefix = {arXiv}
}

@article{Guo:2004dt,
  title = {Heavy quarkonium {$\pi^+ \pi^-$} transitions and a possible $b \bar{b} q \bar{q}$ state},
  author = {Guo, F.-K. and Shen, P.-N. and Chiang, H.-C. and Ping, R.-G.},
  year = 2005,
  journal = {Nucl. Phys. A},
  volume = {761},
  eprint = {hep-ph/0410204},
  pages = {269--282},
  doi = {10.1016/j.nuclphysa.2005.07.019},
  archiveprefix = {arXiv}
}

@article{Pelaez:2024uav,
    author = "Pel{\'a}ez, J. R. and Rab{\'a}n, P. and de Elvira, J. Ruiz",
    title = "{Global parametrizations of $\pi\pi$ scattering with dispersive constraints: Beyond the S0 wave}",
    eprint = "2412.15327",
    archivePrefix = "arXiv",
    primaryClass = "hep-ph",
    reportNumber = "IPARCOS-UCM-24-062",
    doi = "10.1103/PhysRevD.111.074003",
    journal = "Phys. Rev. D",
    volume = "111",
    number = "7",
    pages = "074003",
    year = "2025"
}

@article{Moussallam:1999aq,
    author = "Moussallam, Bachir",
    title = "{$N_f$ dependence of the quark condensate from a chiral sum rule}",
    eprint = "hep-ph/9909292",
    archivePrefix = "arXiv",
    reportNumber = "IPNO-DR-99-22",
    doi = "10.1007/s100520050738",
    journal = "Eur. Phys. J. C",
    volume = "14",
    pages = "111--122",
    year = "2000"
}

@article{Danilkin:2025kyo,
    author = "Danilkin, Igor and Deineka, Oleksandra and Passemar, Emilie and Vanderhaeghen, Marc",
    title = "{Coupled-channel Omn{\`e}s matrix for the $D$-wave isoscalar $\pi\pi/K\bar{K}$ system and its application to $J/\psi \to \pi^{0}\pi^{0}\gamma$, $K_{S}K_{S}\gamma$}",
    eprint = "2512.23669",
    archivePrefix = "arXiv",
    primaryClass = "hep-ph",
    doi = "10.1016/j.physletb.2026.140327",
    journal = "Phys. Lett. B",
    volume = "875",
    pages = "140327",
    year = "2026"
}

@article{James:1975dr,
	author = "James, F. and Roos, M.",
	title = "{Minuit: A System for Function Minimization and Analysis of the Parameter Errors and Correlations}",
	reportNumber = "CERN-DD-75-20",
	doi = "10.1016/0010-4655(75)90039-9",
	journal = "Comput. Phys. Commun.",
	volume = "10",
	pages = "343--367",
	year = "1975"
}

@misc{iminuit,
	author       = {Hans Dembinski and Piti Ongmongkolkul and others},
	title        = {scikit-hep/iminuit},
	month        = {Nov},
	year         = {2025},
	publisher    = {Zenodo},
	version      = {v2.32.0},
	doi          = {10.5281/zenodo.17565861},
}

@article{ParticleDataGroup:2024cfk,
	author = "Navas, S. and others",
	collaboration = "Particle Data Group",
	title = "{Review of particle physics}",
	doi = "10.1103/PhysRevD.110.030001",
	journal = "Phys. Rev. D",
	volume = "110",
	number = "3",
	pages = "030001",
	year = "2024"
}

@article{Pelaez:2020gnd,
    author = "Pel{\'a}ez, Jos{\'e} R. and Rodas, Arkaitz",
    title = "{Dispersive $\pi K \to \pi K$ and $\pi \pi \to K\bar{K}$ amplitudes from scattering data, threshold parameters, and the lightest strange resonance $\kappa$ or $K_{0}^{*}(700)$ }",
    eprint = "2010.11222",
    archivePrefix = "arXiv",
    primaryClass = "hep-ph",
    reportNumber = "JLAB-THY-20-3276",
    doi = "10.1016/j.physrep.2022.03.004",
    journal = "Phys. Rept.",
    volume = "969",
    pages = "1--126",
    year = "2022"
}

@article{Hoferichter:2015hva,
    author = "Hoferichter, Martin and Ruiz de Elvira, Jacobo and Kubis, Bastian and Mei{\ss}ner, Ulf-G.",
    title = "{Roy-Steiner-equation analysis of pion-nucleon scattering}",
    eprint = "1510.06039",
    archivePrefix = "arXiv",
    primaryClass = "hep-ph",
    reportNumber = "INT-PUB-15-050",
    doi = "10.1016/j.physrep.2016.02.002",
    journal = "Phys. Rept.",
    volume = "625",
    pages = "1--88",
    year = "2016"
}

@misc{Cao:2025dkv,
    author = "Cao, Xiong-Hui and Guo, Feng-Kun and Li, Qu-Zhi and Wu, Bo-Wen and Yao, De-Liang",
    title = "{Gravitational form factors of pions, kaons and nucleons from dispersion relations}",
    eprint = "2507.05375",
    archivePrefix = "arXiv",
    primaryClass = "hep-ph",
    doi = "10.1140/epjs/s11734-025-02025-9",
    month = "7",
    year = "2025"
}

@article{Matuschek:2020gqe,
    author = "Matuschek, Inka and Baru, Vadim and Guo, Feng-Kun and Hanhart, Christoph",
    title = "{On the nature of near-threshold bound and virtual states}",
    eprint = "2007.05329",
    archivePrefix = "arXiv",
    primaryClass = "hep-ph",
    doi = "10.1140/epja/s10050-021-00413-y",
    journal = "Eur. Phys. J. A",
    volume = "57",
    number = "3",
    pages = "101",
    year = "2021"
}

\end{document}